\documentclass[10pt,twocolumn,letterpaper]{article}

\usepackage{cvpr}              

\usepackage{graphicx}
\usepackage{amsmath}
\usepackage{amssymb}
\usepackage[accsupp]{axessibility}
\usepackage{booktabs}
\usepackage{multirow}
\usepackage{xcolor}
\usepackage{colortbl}
\usepackage{arydshln}

\definecolor{lightblue}{HTML}{D3E6F1}
\definecolor{bestresult}{HTML}{FFCCC9}

\definecolor{cvprblue}{rgb}{0.21,0.49,0.74}
\usepackage[pagebackref,breaklinks,colorlinks,allcolors=cvprblue]{hyperref}

\def\paperID{46343} 
\def\confName{CVPR}
\def\confYear{2026}

\title{RelativeFlow: Taming Medical Image Denoising Learning with Noisy Reference}

\author{
Yuxin Liu$^{1}$ \quad
Yiqing Dong$^{4}$ \quad
Wenxue Yu$^{1}$ \quad
Zhan Wu$^{1}$ \quad
Rongjun Ge$^{3}$ \quad
Yang Chen$^{1}$\thanks{Corresponding authors.} \quad
Yuting He$^{2}$\footnotemark[1] \\
\\
$^{1}$School of Computer Science and Engineering, Southeast University, China \\
$^{2}$Department of Biomedical Engineering, Case Western Reserve University, USA \\
$^{3}$School of Instrument Science and Engineering, Southeast University, China \\
$^{4}$Department of Optical Sciences, University of Arizona, USA\\
}

\begin{document}
\maketitle
\begin{abstract}
Medical image denoising (MID) lacks absolutely clean images for supervision, leading to a noisy reference problem that fundamentally limits denoising performance. 
Existing simulated-supervised discriminative learning (SimSDL) and simulated-supervised generative learning (SimSGL) treat noisy references as clean targets, causing suboptimal convergence or reference-biased learning, while self-supervised learning (SSL) imposes restrictive noise assumptions that are seldom satisfied in realistic MID scenarios. 
We propose \textbf{RelativeFlow}, a flow matching framework that learns from heterogeneous noisy references and drives inputs from arbitrary quality levels toward a unified high-quality target. 
RelativeFlow reformulates flow matching by decomposing the absolute noise-to-clean mapping into relative noisier-to-noisy mappings, and realizes this formulation through two key components: 1) consistent transport (CoT), a displacement map that constrains relative flows to be components of and progressively compose a unified absolute flow, and 2) simulation-based velocity field (SVF), which constructs a learnable velocity field using modality-specific degradation operators to support different medical imaging modalities. 
Extensive experiments on Computed Tomography (CT) and Magnetic Resonance (MR) denoising demonstrate that RelativeFlow significantly outperforms existing methods, taming MID with noisy references.
\end{abstract}

\section{Introduction}
\label{sec:Introduction}

\begin{figure}[h]
    \centering
    \includegraphics[width=\linewidth]{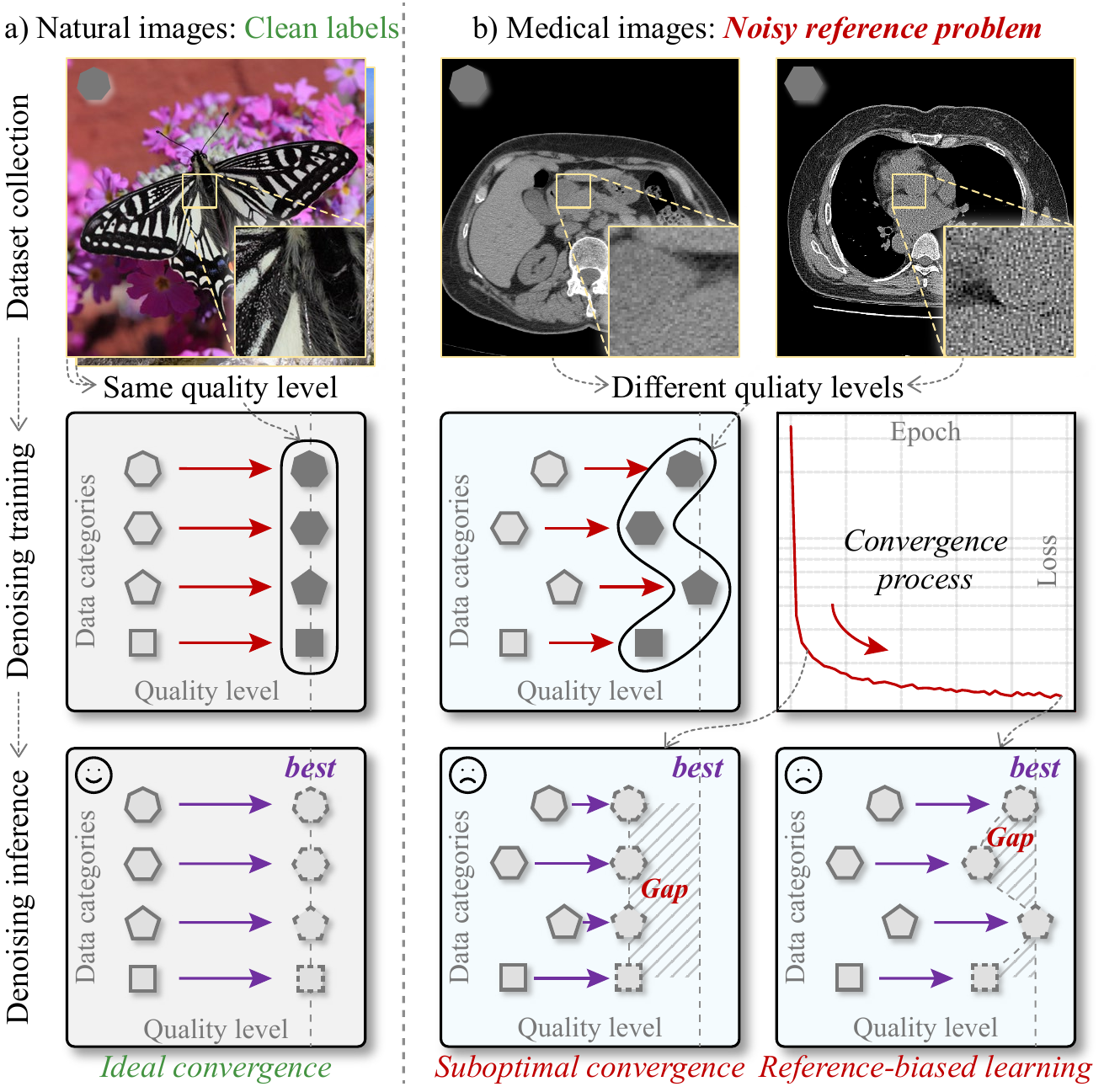}
    \caption{\textbf{Noisy reference problem}: Varying reference quality across categories causes medical image denoising models to learn category-specific mappings, limiting denoising performance.}
    \label{fig:problem}
\end{figure}

Medical image denoising  (MID) suffers from the \textit{noisy reference problem}, which fundamentally limits denoising performance. 
Natural image denoising (NID) typically collects absolutely high-quality images as clean labels for model training~\cite{zhang2017dncnn,zhang2018ffdnet,liang2021swinir}, where all labels share the same quality level (Fig.~\ref{fig:problem} a).
However, MID is constrained by device specifications and acquisition settings~\cite{soylu2022calibrating} in modalities such as Computed Tomography (CT) and Magnetic Resonance (MR)~\cite{zeng2015simple,gudbjartsson1995rician}, providing only relatively high-quality images as noisy references for training, whose quality varies across heterogeneous acquisition protocols and scanner configurations~\cite{deep2021harmonization,quantification2013heterogeneity} (Fig.~\ref{fig:problem} b).
As neural networks learn the noisy-to-clean mapping, NID models uniformly denoise images acquired under different conditions to the best quality, while MID models are trapped in either of two failure modes:
\textit{1) suboptimal convergence}, where all inputs are mapped to a uniform but suboptimal quality level when the model underfits~\cite{song2020learning}, and
\textit{2) reference-biased learning}, where each acquisition condition is mapped to its own reference quality level when the model converges~\cite{song2020learning,arpit2017closer}.
As a result, the model capability is fundamentally limited by noisy references, failing to denoise all images to a consistent high quality.

Current image denoising learning paradigms fail to tame MID with noisy reference problem.
\textit{1) Simulated-Supervised Discriminative Learning (SimSDL)}~\cite{chen2017redcnn,yang2018wganvgg,li2020sacnn,benou2016denoising,gregory2021hydranet,jiang2023armnet}, the most prevalent deep learning approach for MID, suffers from the noisy reference problem.
Specifically, it simulates degraded inputs $\{\tilde{\mathbf{x}}_r\}$ from noisy references $\{\mathbf{x}_r\}$ to learn $f_\theta: \tilde{\mathbf{X}}_r \rightarrow \mathbf{X}_r$, which constitutes an ill-posed absolute noise-to-clean mapping.
\textit{2) Self-supervised learning (SSL)}~\cite{niu2022noise2sim,jing2022training,fadnavis2020patch2self,xu2021deformed2self,tian2022sdndti} constructs and learns relative noisier-to-noisy mapping $f_\theta: \mathbf{X}_r' \rightarrow \mathbf{X}_r$ within noisy references, yet its restrictive assumptions (e.g., independent noise) hinder application in MID.
\textit{3) Simulated-Supervised Generative Learning (SimSGL)}~\cite{gao2023corediff,liao2024idpm,xiang2023ddm2,song2021solving} adapts generative methods to MID via medical image simulation, yet still face this problem by naively treating noisy references as generation targets $\mathbf{x}_1$, which also constitutes an ill-posed absolute noise-to-clean mapping $f_\theta: \mathbf{X}_0 \rightarrow \mathbf{X}_1$.

\begin{figure}[t]
    \centering
    \includegraphics[width=\linewidth]{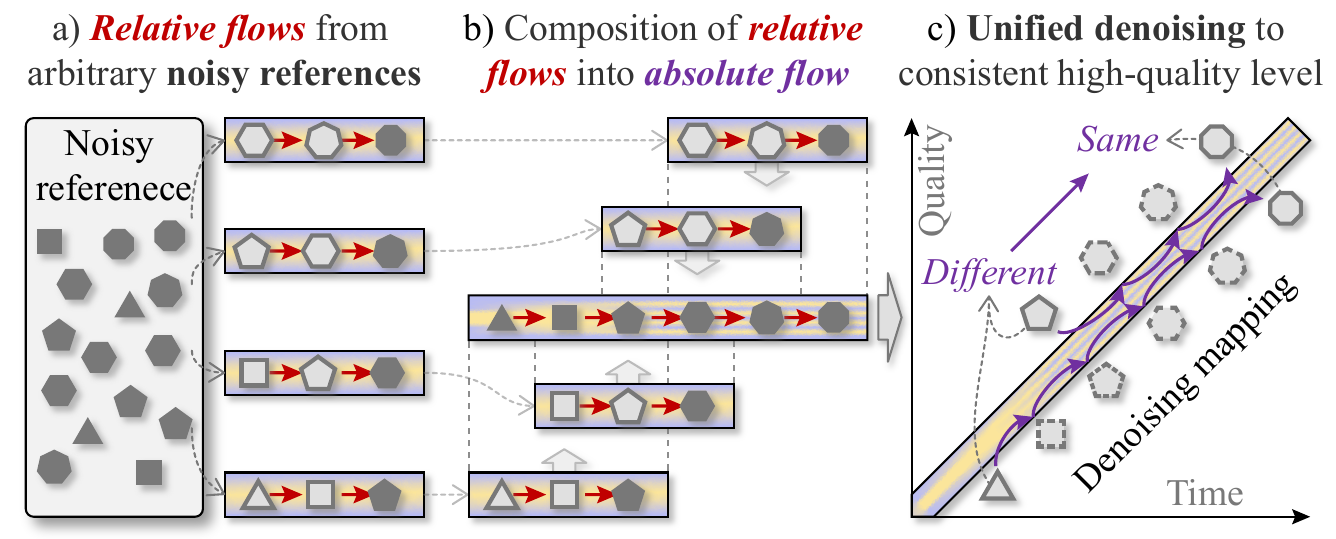}
    \caption{\textbf{Motivation}: Composing relative denoising flows from arbitrary noisy references into an absolute denoising flow transports images with different quality levels to a unified high-quality level, breaking reference bias.}
    \label{fig:motivation}
\end{figure}

\textit{Motivation:} Relative noisier-to-noisy flows $\mathbf{X}_{r} \rightarrow \tilde{\mathbf{X}}_{r}$ are components of and progressively compose a unified absolute noise-to-clean flow $\mathbf{X}_0 \rightarrow \mathbf{X}_1$.
Suppose we can construct relative denoising mapping for arbitrary noisy references $\{\mathbf{x}_r^i\}_{i=1}^N$ to form a unified absolute denoising mapping (Fig.~\ref{fig:motivation} a, b), this mapping will transform noisy images from different quality to a unified high-quality level $\mathbf{x}_1$, breaking through reference bias (Fig.~\ref{fig:motivation} c).
Fortunately, generative methods such as \textit{flow matching}~\cite{lipman2022flow} provide this capability by learning the absolute flow $\psi: \mathbf{X}_0 \rightarrow \mathbf{X}_1$ from relative local flows $u(\mathbf{x}_t|\mathbf{x}_{t+\delta}): \mathbb{R}^d \rightarrow \mathbb{R}^d$ that characterize incremental denoising steps.
However, two fundamental questions remain for practical implementation: 
\textbf{Q1.} How to constrain relative flows $\{\psi^i\}_{i=1}^N$ constructed from different noisy references $\{\mathbf{x}_r^i\}_{i=1}^N$ to a unified flow $\psi$?
\textbf{Q2.} How to leverage medical image simulation methods to construct velocity fields for model learning?

\textit{For the first time}, we formulate the \textbf{RelativeFlow} framework based on flow matching to tame MID with noisy references and apply it across two medical imaging modalities (CT and MR) via two key innovations:
\begin{enumerate}
    \item \textit{For Q1}, we provide Consistent Transport (CoT) displacement map that enables relative flows to be \textit{components of} and \textit{progressively compose} the absolute flow, constraining all individual flows $\{\psi^i\}_{i=1}^N$ to a unified consistent flow $\psi$ across arbitrary quality levels.
    \item \textit{For Q2}, we propose a Simulation-based Velocity Field (SVF) based on generalizable medical image degradation processes, enabling RelativeFlow for different MID tasks across imaging modalities.
\end{enumerate}
Our implementation, including training and evaluation code for both CT and MR denoising, is publicly available at \url{https://github.com/Deliver0/RelativeFlow}.

\section{Related Works}
\label{sec:Related Works}

\textbf{1) Simulated-Supervised Discriminative Learning} (SimSDL) trains discriminative models by simulating degraded inputs from reference images to learn direct denoising mappings.
In NID, discriminative networks are trained on synthetic noise-corrupted images to learn clean label mappings, addressing Gaussian noise~\cite{zhang2017dncnn,tai2017memnet,ronneberger2015unet,yue2019vdn,zhang2018ffdnet}, spatially-variant noise~\cite{zhang2018ffdnet,liang2021swinir}, and signal-dependent noise in real photographs~\cite{guo2019cbdnet,kim2019grdn,zuo2022cfnet,liang2021swinir}.
In MID, discriminative methods simulate modality-specific noise distributions: Poisson-Gaussian noise from X-ray photon statistics for CT denoising~\cite{chen2017redcnn,yang2018wganvgg,li2020sacnn,zeng2015simple} and Rician noise from magnitude signal reconstruction for MR denoising~\cite{benou2016denoising,gregory2021hydranet,jiang2023armnet,gudbjartsson1995rician}.
Unlike NID with absolutely clean references, MID only accesses noisy references whose quality varies across acquisition conditions, leading to reference-biased learning.

\textbf{2) Self-Supervised Learning} (SSL) constructs relative noisier-to-noisy mappings within noisy data itself to avoid dependence on clean references.
In NID, self-supervised methods learn from noisy image pairs~\cite{noise2noise}, exploit blind-spot networks for single noisy images~\cite{batson2019noise2self,noise2void,broaddus2020structn2v}, and leverage spatial correlations for independent noise removal~\cite{noise2noise,batson2019noise2self}.
In MID, self-supervised approaches exploit similarity-based nonlocal correlations for CT denoising~\cite{niu2022noise2sim,jing2022training} and utilize intrinsic repetitions in diffusion-weighted acquisitions for MR denoising~\cite{fadnavis2020patch2self,xu2021deformed2self,tian2022sdndti}.
Unlike NID with independent additive noise, MID faces physics-based signal-dependent noise (Poisson-Gaussian in CT, Rician in MR) with spatial correlations, violating SSL independence assumptions.

\textbf{3) Simulated-Supervised Generative Learning} (SimSGL) trains generative models by learning reverse processes from noise to reference images through score or velocity field prediction.
In NID, denoising diffusion probabilistic models~\cite{ho2020ddpm}, score-based stochastic differential equations~\cite{song2020sde}, and flow matching~\cite{lipman2022flow} learn to iteratively denoise by predicting noise, score, or velocity fields that reverse degradation from clean to noisy images.
In MID, CT denoising methods employ error-modulated diffusion with mean-preserving degradation operators~\cite{gao2023corediff} or iterative partial diffusion~\cite{liao2024idpm}, while MR denoising integrates statistic-based priors into conditional diffusion generation~\cite{xiang2023ddm2} or leverages score-based posterior sampling for inverse problems~\cite{song2021solving}.
Unlike NID with clean target distributions, MID naively treats noisy references as generation targets, also causing reference-biased learning.

\section{Methodology}
\label{sec:Methodology}

\begin{figure}[h]
    \centering
    \includegraphics[width=\linewidth]{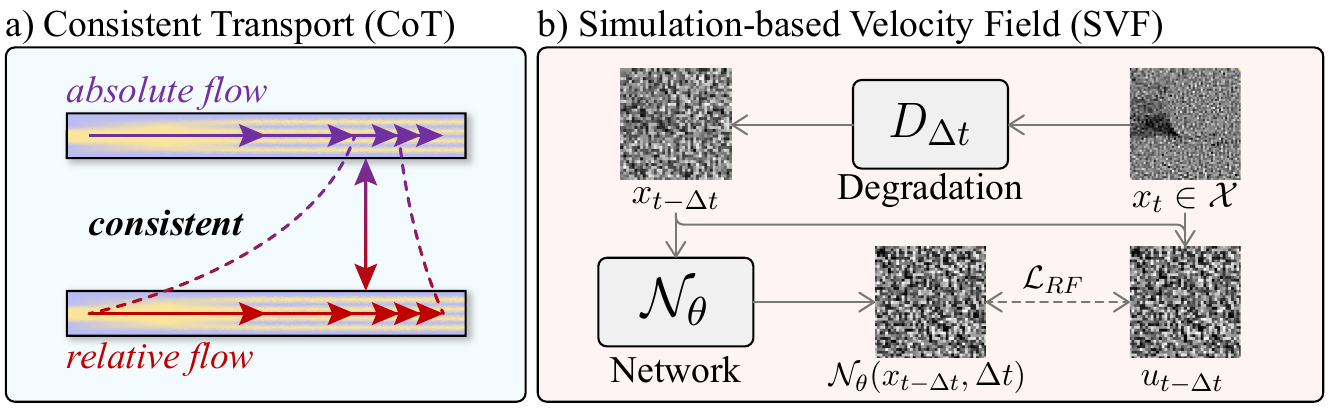}
    \caption{\textbf{Overview of RelativeFlow framework}: Our method learns relative flows from noisy references at varying quality levels through Consistent Transport (CoT) and Simulation-based Velocity Field (SVF), enabling unified denoising across different quality levels.}
    \label{fig:method}
\end{figure}

Our RelativeFlow framework addresses the noisy reference problem via two key innovations: Consistent Transport (CoT, Sec.~\ref{sec:consistent_transport}) that enables relative flows to be components of and progressively compose the absolute flow, constraining flows constructed from different noisy references to a unified consistent flow, and Simulation-based Velocity Field (SVF, Sec.~\ref{sec:simulation_velocity_field}) based on generalizable medical image degradation processes that enables RelativeFlow across different imaging modalities.

\subsection{Problem Formulation}
\label{sec:problem_formulation}

Our RelativeFlow reformulates medical image denoising as learning a continuous flow along a latent quality--time axis driven by a velocity field. Let $\mathcal{X} = \{x_t\}$ denote noisy reference images acquired under heterogeneous protocols, parameterized by $t \in (0,+\infty)$, where $t=0$ denotes the noise endpoint and larger $t$ corresponds to progressively cleaner states approaching a theoretical clean limit as $t \to +\infty$. This induces a family of marginal distributions $\{p_t\}$ forming an absolute denoising path from $p_0$ to $p_{+\infty}$. In practice, neither the clean endpoint $p_{+\infty}$ nor the absolute times $\{t_i\}$ are observed; only noisy references at unknown quality levels $t_i>0$ are available.

To obtain supervision we use a modality-specific medical image degradation operator $D_{\Delta t}$ (Sec.~\ref{sec:simulation_velocity_field}). For any noisy reference $x_t \sim p_t$ at arbitrary and unknown $t>0$, applying $D_{\Delta t}$ produces a more degraded sample $x_{t-\Delta t} = D_{\Delta t}(x_t)$ at an earlier quality level $t-\Delta t$ for any step size $\Delta t>0$. The pair $(x_{t-\Delta t}, x_t)$ defines a local relative denoising step along the underlying absolute flow.

Our goal is to learn a denoising flow $\psi : (0,+\infty) \times \mathbb{R}^d \to \mathbb{R}^d$ whose restriction on any interval $[t-\Delta t, t]$ matches these relative steps and transports arbitrary noisy inputs toward higher quality levels. To achieve this, we construct a learning objective $\mathcal{L}_{RF}$ that trains a neural network $\mathcal{N}_\theta$ to predict the CoT-consistent velocity field from simulated degraded pairs, with the explicit formulation derived in Sec.~\ref{sec:simulation_velocity_field}.

\subsection{Consistent Transport}
\label{sec:consistent_transport}

CoT establishes the displacement map between arbitrary distributions $p_{t_i}$ and $p_{t_j}$, enabling relative flows $\psi_{t_i \to t_j}$ to be \textit{components of} and \textit{progressively compose} the absolute flow $\psi_{t_0 \to t_{+\infty}}$, with the probability path as:
\begin{equation}
p_t = \lambda \, p_{t_i} + (1 - \lambda) \, p_{t_j}, \quad \lambda = \frac{e^{-t} - e^{-t_j}}{e^{-t_i} - e^{-t_j}}
\label{eq:relative_flow}
\end{equation}
where $0 < t_i < t < t_j$ defines the probability path of relative flows between arbitrary quality ranges.
This yields the absolute flow $p_t = e^{-t} p_0 + (1-e^{-t}) p_{+\infty}$ for $t \in (0, +\infty)$, with $p_0$ and $p_{+\infty}$ obtained in the limits $t \to 0$ and $t \to +\infty$.

\textbf{Why are relative flows \textit{components of} absolute flow?}
\begin{figure}[h]
    \centering
    \includegraphics[width=\linewidth]{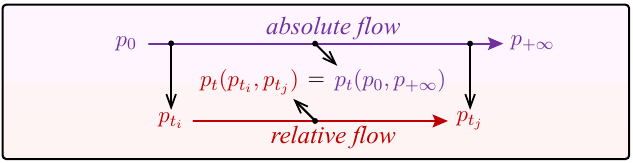}
    \caption{Relative flows are components of absolute flow.}
    \label{fig:cot_proof1}
\end{figure}
As shown in Fig.~\ref{fig:cot_proof1}, CoT defines the probability path via linear interpolation in exponential time space with weight $\lambda = \frac{e^{-t} - e^{-t_j}}{e^{-t_i} - e^{-t_j}}$, ensuring that nested interpolations preserve transitivity—any intermediate distribution remains expressible as a direct interpolation between the original endpoints.
To prove this, consider arbitrary quality levels $0 < t_i < t < t_j < +\infty$ where both $p_{t_i} = e^{-t_i} p_0 + (1-e^{-t_i}) p_{+\infty}$ and $p_{t_j} = e^{-t_j} p_0 + (1-e^{-t_j}) p_{+\infty}$ lie on the absolute flow path.
Substituting these into the relative path definition (Eq.~\eqref{eq:relative_flow}) yields $p_t = e^{-t} p_0 + (1-e^{-t}) p_{+\infty}$, exactly matching the absolute flow formula (detailed derivation in \textit{Supplementary}).
Therefore, any distribution in a relative flow is a component of the absolute flow, ensuring that training on diverse noisy references $\{x^i\}_{i=1}^N$ at varying quality levels $\{t_i\}_{i=1}^N$ constructs local relative flows that compose into the unified absolute flow.

\textbf{Why do relative flows \textit{progressively compose} absolute flow?}
\begin{figure}[h]
    \centering
    \includegraphics[width=\linewidth]{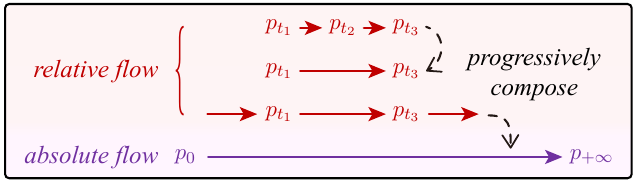}
    \caption{Relative flows progressively compose absolute flow.}
    \label{fig:cot_proof2}
\end{figure}
As shown in Fig.~\ref{fig:cot_proof2}, CoT guarantees that composing consecutive flows yields absolute flow, enabling practical flow composition during training.
Specifically, for any three quality levels $0 < t_1 < t_2 < t_3$, applying flow $\psi_{t_1 \to t_2}$ followed by $\psi_{t_2 \to t_3}$ is equivalent to directly applying $\psi_{t_1 \to t_3}$, i.e., $\psi_{t_1 \to t_3} = \psi_{t_2 \to t_3} \circ \psi_{t_1 \to t_2}$.
To verify this property, we express distributions at intermediate quality level $t_2$ using both the flow from $t_1$ and the flow to $t_3$.
For flow $\psi_{t_1 \to t_2}$, the probability path gives $p_t = \lambda_{12} p_{t_1} + (1-\lambda_{12}) p_{t_2}$.
Similarly, for flow $\psi_{t_2 \to t_3}$, we have $p_t = \lambda_{23} p_{t_2} + (1-\lambda_{23}) p_{t_3}$.
Eliminating $p_{t_2}$ yields $p_t = \lambda_{13} p_{t_1} + (1-\lambda_{13}) p_{t_3}$ (detailed derivation in \textit{Supplementary}), which exactly matches the CoT probability path definition (Eq.~\eqref{eq:relative_flow}) for the direct flow from $t_1$ to $t_3$.
This confirms that composing two consecutive relative flows produces the same probability path as the direct relative flow, establishing the composition property.
By this property, any sequence of relative flows $\psi_{t_1 \to t_2}, \psi_{t_2 \to t_3}, \ldots, \psi_{t_{k-1} \to t_k}$ from noisy references at heterogeneous quality levels $\{t_i\}_{i=1}^N$ composes into a single flow $\psi_{t_1 \to t_k} = \psi_{t_{k-1} \to t_k} \circ \cdots \circ \psi_{t_1 \to t_2}$.
During training, the model observes diverse relative flows spanning different quality ranges, progressively learning the velocity field across the entire quality spectrum $(0, +\infty)$ to construct the complete continuous absolute flow $\psi_{0 \to +\infty}$.
Consequently, the trained model breaks through reference bias by denoising images different quality levels toward a unified optimal quality level $t \to +\infty$, rather than being constrained to acquisition-specific reference quality.

\begin{table}[t]
\centering
\caption{Training and Sampling Algorithms for RelativeFlow}
\label{alg:algorithms}
\begin{tabular*}{\linewidth}{@{\extracolsep{\fill}}l@{}}
\toprule
\textbf{Algorithm 1} Training \\
\textbf{Input:} Epochs $E$, Dataset $\mathcal{X}$, $\Delta t_{\min}$, $\Delta t_{\max}$, $\alpha < 1$ \\
\textbf{Output:} Model $\mathcal{N}_\theta$ \\
\midrule
1: \textbf{for} $e \in \{1, \ldots, E\}$ \textbf{do} \\
2: \quad \textbf{while} not converged \textbf{do} \\
3: \quad\quad Sample $x_t \sim \mathcal{X}$ \\
4: \quad\quad Sample $\Delta t \sim \text{Uniform}[\Delta t_{\min}, \Delta t_{\max}]$ \\
5: \quad\quad Generate $x_{t-\Delta t} \leftarrow D_{\Delta t}(x_t)$ \\
6: \quad\quad Target $u \leftarrow \dfrac{x_t - x_{t-\Delta t}}{e^{\Delta t}-1}$ \\
7: \quad\quad Take gradient descent step on \\
\quad\quad\quad\quad\quad\quad\quad $\nabla_\theta \left\|\mathcal{N}_\theta(x_{t-\Delta t}, \Delta t) - u\right\|^2$ \\
8: \quad \textbf{end while} \\
9: \quad Update $\Delta t_{\min} \leftarrow \alpha \cdot \Delta t_{\min}$, $\Delta t_{\max} \leftarrow \Delta t_{\max} / \alpha$ \\
10: \textbf{end for} \\
\toprule
\textbf{Algorithm 2} Sampling \\
\textbf{Input:} Noisy image $x_{noisy}$, Model $\mathcal{N}_\theta$, $\{\Delta t_i\}$ \\
\textbf{Output:} Denoised image $x_{clean}$ \\
\midrule
1: Initialize $x \leftarrow x_{noisy}$ \\
2: \textbf{for} $\Delta t \in \{\Delta t_i\}$ \textbf{do} \\
3: \quad Predict $u \leftarrow \mathcal{N}_\theta(x, \Delta t)$ \\
4: \quad Update $x \leftarrow x + \Delta t \cdot u$ \\
5: \textbf{end for} \\
6: Set $x_{clean} \leftarrow x$ \\
7: \textbf{return} $x_{clean}$ \\
\bottomrule
\end{tabular*}
\end{table}

\subsection{Simulation-based Velocity Field}
\label{sec:simulation_velocity_field}

Given a noisy reference $x_t$ at arbitrary quality level $t$ and a degradation operator $D_{\Delta t}$ that generates a noisier sample $x_{t-\Delta t} = D_{\Delta t}(x_t)$, we define the training loss for velocity field prediction as:
\begin{equation}
\mathcal{L}_{RF} = \mathbb{E}_{x_t,\,\Delta t}\!\left[\left\|\mathcal{N}_\theta\!\left(D_{\Delta t}(x_t),\, \Delta t\right) - \frac{x_t - D_{\Delta t}(x_t)}{e^{\Delta t}-1}\right\|_2^2\right].
\label{eq:training_loss}
\end{equation}

The practical implementation follows Alg.~\ref{alg:algorithms}, where training uses simulated relative degradations around each noisy reference to supervise the velocity field, and sampling iteratively integrates the predicted velocity.

\begin{table*}[ht]
\centering
\caption{\textbf{Quantitative comparison}: 
Our RelativeFlow achieves the best performance compared with 10 methods across 4 metrics on a) CT and b) MR denoising tasks. All methods use unified settings; see the \textit{Supplementary} for details. 
\colorbox{bestresult}{Red} and \colorbox{lightblue}{blue} backgrounds indicate best and second-best results.
$\dagger$ and * represent NID and MID methods.
Images are normalized to [0,1] range for computing PSNR, SSIM, and RMSE, and to 3-channel [0,255] format for LPIPS computation. SSIM, RMSE, and LPIPS values are reported as their raw scores multiplied by 100 for clearer comparison. 
}
\label{tab:combined_quant}
\resizebox{\textwidth}{!}{%
\begin{tabular}{c|l|cccc||l|cccc}
& \multicolumn{5}{c||}{\textbf{a) CT denoising evaluation}} & \multicolumn{5}{c}{\textbf{b) MR denoising evaluation}} \\
Type & Method & PSNR$\uparrow$ & SSIM$\uparrow$ & RMSE$\downarrow$ & LPIPS$\downarrow$ & Method & PSNR$\uparrow$ & SSIM$\uparrow$ & RMSE$\downarrow$ & LPIPS$\downarrow$ \\
\hline
\multirow{2}{*}{SimSDL} & SwinIR$^\dagger$~\cite{liang2021swinir} & 41.78 & \cellcolor{lightblue}93.12 & 1.13 & 4.62 & SwinIR$^\dagger$~\cite{liang2021swinir} & 34.35 & \cellcolor{lightblue}92.38 & \cellcolor{lightblue}1.55 & 3.68 \\
& RED-CNN$^*$~\cite{chen2017redcnn} & \cellcolor{lightblue}43.12 & 92.15 & \cellcolor{lightblue}0.74 & 5.18 & CNN-DMRI$^*$~\cite{tripathi2020cnn} & 32.03 & 80.21 & 1.69 & 4.91 \\
\hdashline[1.2px/2px]
\multirow{2}{*}{SSL} & Noise2Self$^\dagger$~\cite{batson2019noise2self} & 36.45 & 81.55 & 1.52 & 19.85 & Noise2Self$^\dagger$~\cite{batson2019noise2self} & 30.48 & 88.72 & 3.45 & 3.82 \\
& Noise2Sim$^*$~\cite{niu2022noise2sim} & 36.35 & 88.71 & 1.52 & 9.18 & Deformed2Self$^*$~\cite{xu2021deformed2self} & 27.28 & 79.91 & 5.29 & 5.51 \\
\hdashline[1.2px/2px]
\multirow{4}{*}{SimSGL} & DDIM$^\dagger$~\cite{song2020ddim} & 35.71 & 89.58 & 1.64 & 11.21 & DDIM$^\dagger$~\cite{song2020ddim} & 29.52 & 70.05 & 4.53 & 5.85 \\
& Flow Matching$^\dagger$~\cite{lipman2022flow} & 42.05 & 92.88 & 0.83 & \cellcolor{lightblue}4.51 & Flow Matching$^\dagger$~\cite{lipman2022flow} & \cellcolor{lightblue}35.18 & 85.42 & 1.81 & \cellcolor{lightblue}3.49 \\
& IPDM$^*$~\cite{liao2024idpm} & 31.31 & 92.42 & 2.82 & 5.97 & R2D2+$^*$~\cite{chung2022r2d2} & 31.96 & 71.45 & 2.57 & 4.46 \\
& \textbf{RelativeFlow$^*$ (Ours)} & \cellcolor{bestresult}\textbf{43.38} & \cellcolor{bestresult}\textbf{93.59} & \cellcolor{bestresult}\textbf{0.70} & \cellcolor{bestresult}\textbf{4.38} & \textbf{RelativeFlow$^*$ (Ours)} & \cellcolor{bestresult}\textbf{36.16} & \cellcolor{bestresult}\textbf{93.21} & \cellcolor{bestresult}\textbf{1.49} & \cellcolor{bestresult}\textbf{3.01} \\
\end{tabular}%
}
\end{table*}

\textbf{Why SVF enables simulation-constrained velocity fields with arbitrary noisy references.}
Since medical imaging physics ensures consistent degradation processes across arbitrary quality levels, the degradation operator $D_{\Delta t}$ applies to any noisy reference $x_t$, enabling velocity field construction for relative flow learning.
Following CoT-defined flow matching, let $\psi_t$ denote the flow map and $x_{\infty}$ the clean endpoint. For $x$ drawn from the noise endpoint, the CoT path yields $\psi_t(x) = e^{-t} x + (1-e^{-t}) x_{\infty}$.
Then the velocity field equals the time derivative along this path:
\begin{equation}
u_t\big(\psi_t(x)\,|\,x, x_{\infty}\big) = \frac{d}{dt}\, \psi_t(x) = e^{-t}(x_{\infty}-x).
\label{eq:cot_velocity_field_absolute}
\end{equation}
However, in practice, only the noisy reference $x_t$ and its degraded version $x_{t-\Delta t} = D_{\Delta t}(x_t)$ are available.
Therefore, we need to express $(x, x_{\infty})$ in Eq.~\eqref{eq:cot_velocity_field_absolute} using the observable pair $(x_t, x_{t-\Delta t})$.
Since both lie on the same CoT path, we can eliminate the unobservable endpoints through algebraic manipulation.
Substituting this into the velocity field expression yields
\begin{equation}
u_{t - \Delta t}\big(x_{t-\Delta t},\Delta t\big) = \frac{x_t - x_{t-\Delta t}}{e^{\Delta t}-1},
\end{equation}
which matches the supervision target in Eq.~\eqref{eq:training_loss}. Hence, simulated relative degradations around arbitrary noisy references consistently constrain the velocity field without requiring access to clean targets or absolute time labels.

\section{Experiments}
\label{sec:Experiments}

\subsection{Comparative Experiments}
\label{sec:comparative_experiments}

\subsubsection{Experimental Protocol} 
We conduct comprehensive experiments on CT and MR denoising to evaluate RelativeFlow across different imaging modalities and noise distributions. The following summarizes the evaluation protocol; full implementation details are provided in the \textit{Supplementary}.

\textit{\textbf{1) Materials}}: This study evaluated RelativeFlow on CT and MR denoising across different medical imaging modalities.
\textit{\textbf{a}) CT Denoising:} Evaluation uses CT images from the GBA-LDCT dataset~\cite{gba2024ldct}, which contain quantum noise and electronic noise following Poisson-Gaussian distribution~\cite{zeng2015simple}. Comprising 263 training patients acquired across different manufacturers and scanning protocols (varying tube voltages, tube currents, and reconstruction kernels), the dataset exhibits substantial reference quality variation across samples.
\textit{\textbf{b}) MR Denoising:} Evaluation uses MR images from the IXI dataset~\cite{ixi2012dataset}, which contain thermal noise and dielectric losses following Rician distribution~\cite{gudbjartsson1995rician}. Comprising 497 training subjects from multiple medical centers with scanners from different manufacturers at varying field strengths and pulse sequences, the dataset exhibits heterogeneous noise levels across acquisition sites.
These multi-center, multi-protocol datasets contain images at heterogeneous quality levels, faithfully representing the noisy reference problem in real clinical scenarios.
Specifically, we selected relatively highest-quality images as the testing set and simulated two degradation levels representing mild and severe noise conditions as inputs, where the remaining data served as the training set.

\textit{\textbf{2) Comparisons}}: This study benchmarked RelativeFlow against 10 recent and classic methods across three types: SimSDL~\cite{liang2021swinir,chen2017redcnn,tripathi2020cnn}, SSL~\cite{batson2019noise2self,niu2022noise2sim,xu2021deformed2self}, and SimSGL~\cite{song2020ddim,lipman2022flow,liao2024idpm,chung2022r2d2}. 
For each type, we included methods from both natural image denoising (NID) and medical image denoising (MID). 
We employed 2D U-Net~\cite{ronneberger2015u} as the backbone $\mathcal{N}_\theta$ for SimSDL and SSL methods (except for methods with specific architectural contributions like RED-CNN), and 2D Guided U-Net~\cite{dhariwal2021diffusion} for SimSGL methods with additional time conditioning.
All methods are trained with L2 loss, and SimSGL methods use 3 inference steps.
 
\textit{\textbf{3) Implementation and Evaluation Metrics}}: All tasks were implemented in PyTorch and optimized by Adam with a learning rate of $10^{-4}$. The models were trained for 30 epochs until convergence. Training and testing were conducted on NVIDIA V100 PCIE GPU with 32 GB memory. 
This study used Peak Signal-to-Noise Ratio (PSNR) and Normalized Mean Square Error (RMSE) for pixel-level accuracy assessment, and Structural Similarity Index Measure (SSIM) and Learned Perceptual Image Patch Similarity (LPIPS) for perceptual consistency evaluation. 
Images are normalized to [0,1] range for computing PSNR, SSIM, and RMSE, and to 3-channel [0,255] format for LPIPS computation. For clearer numerical comparison in Tab.~\ref{tab:combined_quant}, SSIM, RMSE, and LPIPS are reported as their raw scores multiplied by 100.

\subsubsection{Quantitative Analysis}
\label{sec:quantitative_analysis}

Our comprehensive evaluation across CT and MR denoising tasks demonstrates RelativeFlow's superior capability in taming medical image denoising with noisy references shown in Tab.~\ref{tab:combined_quant}. RelativeFlow consistently achieves SOTA performance across both medical imaging modalities:

\begin{figure*}[t]
    \centering
    \includegraphics[width=\linewidth]{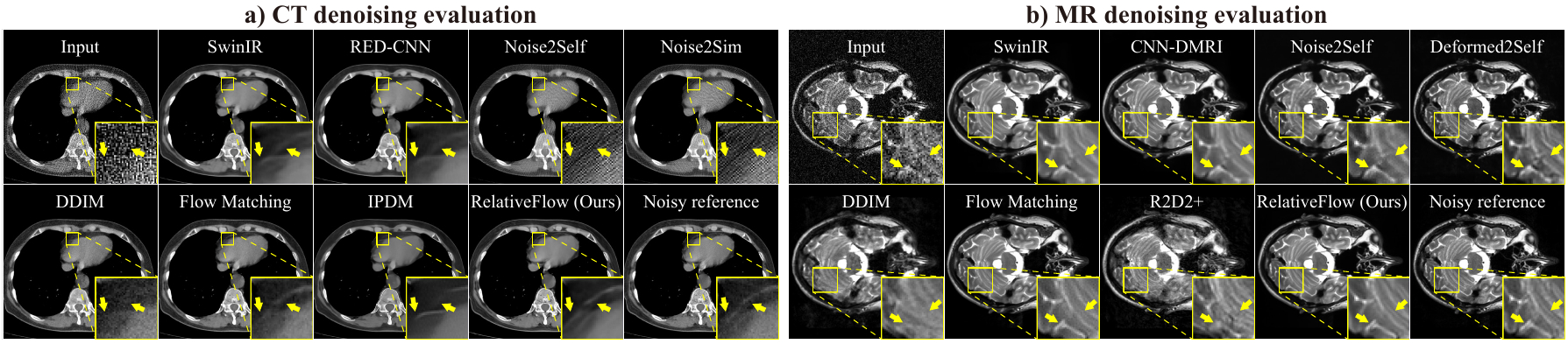}
    \caption{\textbf{Qualitative comparison}: 
    Our RelativeFlow delivers visually superior denoising on a) CT and b) MR images, recovering fine anatomical structures with clearly discernible textures that even surpass those visible in the noisy references.  
    The figure shows visual results for CT (display window [-160, 240] HU) and MR denoising compared against 10 SimSDL, SSL, and SimSGL baseline methods.}
    \label{fig:qualitative_results}
\end{figure*}

\textbf{1) CT denoising.} RelativeFlow achieves SOTA results across all metrics, demonstrating the best pixel-level accuracy and structural consistency throughout the denoising process. 
Our method attains PSNR of 43.38 dB, SSIM of 93.59, RMSE of 0.70, and LPIPS of 4.38, outperforming the best competing results by 0.26 dB in PSNR, 0.47 in SSIM, and by relative reductions of 5.4\% in RMSE and 2.9\% in LPIPS, respectively. 
These gains come from learning relative flows across CT scans with different noise levels, which better capture CT imaging physics and explicitly separate noise and artifacts from underlying anatomical structures. 
SimSDL methods achieve at most 43.12 dB PSNR, while their SSIM stays at or below 93.12 and LPIPS remains no lower than 4.62, reflecting reference-biased learning from noisy targets that leads to over-smoothed or protocol-specific mappings. 
SSL methods yield substantially worse quantitative results (PSNR around 36 dB, SSIM below 88.71, and LPIPS above 9.18), as their independence assumptions break down under signal-dependent, spatially correlated noise. 
Among SimSGL baselines, the strongest Flow Matching variant reaches 42.05 dB PSNR and 4.51 LPIPS but still lags behind RelativeFlow, since treating noisy references as generation targets constrains denoising performance to heterogeneous reference quality levels and prevents consistent transport toward a unified high-quality endpoint.

\textbf{2) MR denoising.} RelativeFlow achieves SOTA performance on MR denoising under Rician noise, providing the best pixel-wise fidelity and structural consistency across all evaluation metrics. 
Our method attains PSNR of 36.16 dB, SSIM of 93.21, RMSE of 1.49, and LPIPS of 3.01, outperforming the best competing results by 0.98 dB in PSNR, 0.83 in SSIM, and by relative reductions of 3.9\% in RMSE and 13.8\% in LPIPS, respectively. 
These gains arise from SVF's physics-informed degradation modeling, which explicitly captures Rician noise characteristics in MR imaging and constrains the learned velocity field for robust denoising. 
SimSDL methods achieve at most 34.35 dB PSNR, with SSIM at or below 92.38 and LPIPS no lower than 3.68, indicating reference-biased learning from heterogeneous noisy targets that induces protocol-dependent mappings. 
SSL methods yield substantially worse quantitative results (PSNR below 30.48 dB, SSIM below 88.72, and LPIPS above 3.82), because their independence assumptions are violated by Rician, signal-dependent and spatially correlated noise. 
Among SimSGL baselines, the strongest Flow Matching variant reaches 35.18 dB PSNR and 3.49 LPIPS but still lags behind RelativeFlow, since treating noisy references as generation targets prevents consistent transport toward a unified high-quality endpoint.
The consistent gains across both modalities, which follow fundamentally different noise distributions, demonstrate the cross-modal generalizability of SVF's physics-based degradation modeling.

\subsubsection{Qualitative Analysis}
\label{sec:qualitative_analysis}

Visual comparisons demonstrate RelativeFlow's capability to reconstruct accurate anatomical details from severely noise-corrupted images, where fine structures in noisy references require considerable effort to observe.

\textbf{CT Denoising.} As shown in Fig.~\ref{fig:qualitative_results} a, our method accurately recovers both structural integrity and fine anatomical detail from heavily degraded CT images, yielding results that surpass even the noisy reference in clarity.
SimSDL methods (SwinIR and RED-CNN) yield anatomically unreliable results, where cardiac boundary regions appear faint and indistinct and structures recovered from noise in other anatomical regions remain inconsistent with the reference, a direct consequence of suboptimal convergence caused by learning from noisy targets.
SSL methods (Noise2Self and Noise2Sim) exhibit heavy residual noise under the severe degradation conditions in our evaluation, as their independence assumptions break down when signal-dependent noise correlations are misinterpreted as structural features.
Among SimSGL baselines, DDIM achieves only limited noise suppression with poor fine-detail recovery. Flow Matching restores structurally plausible anatomy but lacks sufficient sharpness in fine-detail regions. IPDM produces visually crisp outputs that nonetheless exhibit structural distortions inconsistent with the reference, a hallucination risk inherent when generative models treat noisy references as clean generation endpoints.
In contrast, our RelativeFlow produces denoised results that are structurally accurate, visually sharp, and exhibit lower residual noise than the noisy references themselves, exemplifying the method's unique capability to elevate images from diverse quality levels to a consistently high-quality target.

\textbf{MR Denoising.} As shown in Fig.~\ref{fig:qualitative_results} b, our method recovers both indicated fine-detail regions in the brain parenchyma and produces overall sharp and structurally accurate denoised images. SimSDL methods (SwinIR and CNN-DMRI) achieve visually cleaner results with reduced overall noise, but the fine edge details in the indicated anatomical regions are not recovered, a consequence of reference-biased learning that fails to capture subtle tissue boundaries. SSL methods (Noise2Self and Deformed2Self) produce blurry results where fine anatomical details remain unclear, since the signal-dependent and spatially correlated Rician noise violates their independence assumptions. Among SimSGL baselines, DDIM retains residual noise with blurred structures that deviate substantially from the reference, as its generative prior is anchored to the noisy reference distribution rather than the true clean target. Flow Matching achieves moderate noise reduction and successfully recovers one indicated fine-detail region, but residual noise persists and the other region remains unrestored. R2D2+ produces insufficiently sharp results with overall structural patterns that differ substantially from the reference, reflecting the limitation of treating noisy references as clean generation targets.
In contrast, our RelativeFlow recovers both indicated fine-detail regions and yields denoised images that are sharp and structurally accurate throughout, as CoT's clean-limit formulation and SVF's physics-consistent supervision jointly drive transport beyond any individual noisy reference quality.

\subsection{Ablation Studies}
\label{sec:ablation_studies}

\subsubsection{Component Ablations}
We further analyze the individual contributions of CoT and SVF through component ablations. The baseline is flow matching (FM) with standard linear interpolation and na\"ive pairwise supervision. ``+CoT'' replaces the linear interpolation path with the exponential-time interpolation $\lambda = \frac{e^{-t} - e^{-t_j}}{e^{-t_i} - e^{-t_j}}$, while ``+SVF'' replaces na\"ive absolute-noise supervision by constructing $x_{t-\Delta t}$ with the degradation operator $D_{\Delta t}(x_t)$. This design isolates the effects of the transport path and the supervision signal.

Tab.~\ref{tab:component_ablation_results} shows that both components improve denoising performance, and their combination gives the strongest results on both CT and MR. CoT alone brings consistent but modest gains over the FM baseline, improving PSNR from 42.05 to 42.15 on CT and from 35.18 to 35.22 on MR, which indicates that a quality-aware interpolation path already improves transport consistency across noisy references. SVF produces substantially larger gains, especially on MR, where PSNR increases to 35.87, SSIM rises from 85.42 to 91.56, and LPIPS decreases from 3.49 to 3.15. This confirms that physics-consistent relative supervision is critical for separating noise from weak anatomical structures under heterogeneous reference quality. Combining CoT and SVF further improves PSNR to 43.38 on CT and 36.16 on MR, outperforming the FM baseline by 1.33 dB and 0.98 dB, respectively. These results show that CoT and SVF are complementary: CoT provides a more suitable transport path across quality levels, while SVF supplies supervision that better matches the underlying degradation process.

\begin{table}[h]
\centering
\caption{Component ablation of CoT and SVF on CT and MR denoising.}
\label{tab:component_ablation_results}
\resizebox{\columnwidth}{!}{%
\begin{tabular}{l|cccc||cccc}
& \multicolumn{4}{c||}{\textbf{CT denoising}} & \multicolumn{4}{c}{\textbf{MR denoising}} \\
Setting & PSNR$\uparrow$ & SSIM$\uparrow$ & RMSE$\downarrow$ & LPIPS$\downarrow$ & PSNR$\uparrow$ & SSIM$\uparrow$ & RMSE$\downarrow$ & LPIPS$\downarrow$ \\
\hline
FM (Baseline) & 42.05 & 92.88 & 0.83 & 4.51 & 35.18 & 85.42 & 1.81 & 3.49 \\
FM + CoT & 42.15 & 93.07 & 0.79 & 4.49 & 35.22 & 86.45 & 1.79 & 3.48 \\
FM + SVF & 42.52 & 93.34 & 0.72 & 4.41 & 35.87 & 91.56 & 1.68 & 3.15 \\
FM + CoT + SVF (Ours) & \textbf{43.38} & \textbf{93.59} & \textbf{0.70} & \textbf{4.38} & \textbf{36.16} & \textbf{93.21} & \textbf{1.49} & \textbf{3.01} \\
\end{tabular}%
}
\end{table}

\subsubsection{Hyperparameter Ablations}
The decay factor $\alpha$ in training and the time step schedule $\{\Delta t_i\}$ in sampling (Alg.~\ref{alg:algorithms}) are two key hyperparameters in our RelativeFlow. This study conducts an ablation of them on CT and MR denoising tasks to validate their impact on model performance. The results are as follows:

\begin{table}[h]
\centering
\caption{Hyperparameter ablation of $\alpha$ and $\{\Delta t_i\}$ with PSNR as evaluation metric.}
\label{tab:ablation_results}
\resizebox{\columnwidth}{!}{%
\begin{tabular}{c|cccc}
\hline
$\alpha = $ & 0.8 & 0.9 & 1.0 & 1.1 \\
\hline
CT & 43.21 & 43.38 & 40.85 & 36.39 \\
MR & 35.83 & 36.16 & 34.83 & 31.47 \\
\hline
$\{\Delta t_i\} = $ & [0.3,0.15,0.075] & [0.2,0.1,0.05] & [0.1,0.1,0.1] & [0.05,0.1,0.2] \\
\hline
CT & 42.87 & 43.38 & 42.93 & 40.45 \\
MR & 35.92 & 36.16 & 37.89 & 36.54 \\
\hline
\end{tabular}%
}
\end{table}

\begin{figure*}[h]
    \centering
    \includegraphics[width=\linewidth]{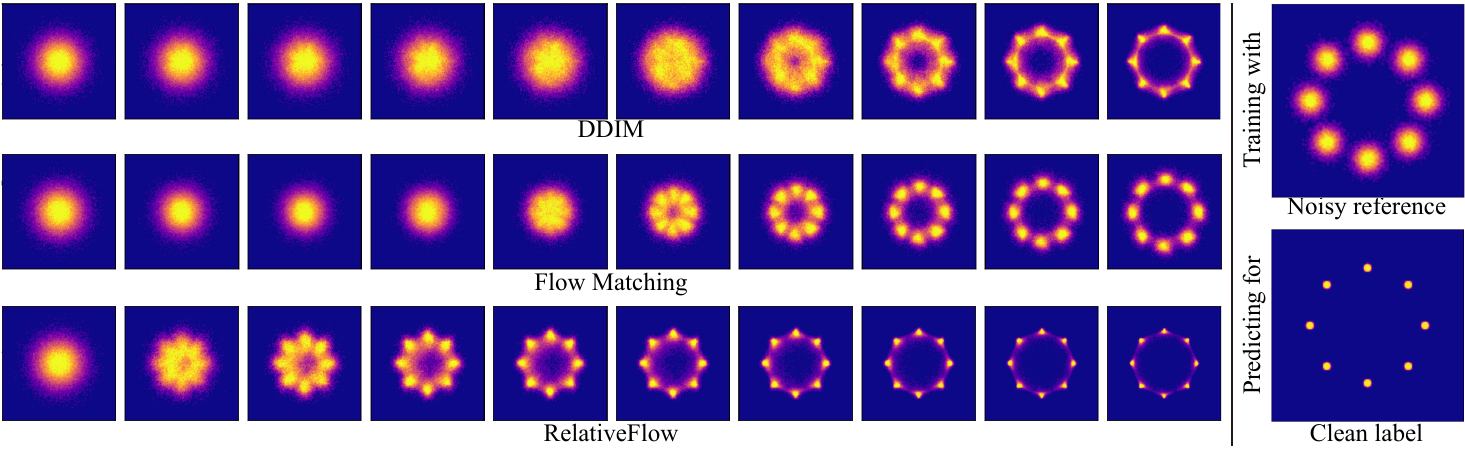}
    \caption{\textbf{Trajectory analysis}: Comparison of transport trajectories learned by DDIM, Flow Matching, and RelativeFlow, training with noisy references (blurred Gaussians) and predicting for clean labels (sharp circular distributions). Our method introduces the ring pattern earlier and generates results closer to clean targets.}
    \label{fig:trajectory_analysis}
\end{figure*}

The decay factor $\alpha$ ($\alpha < 1$) controls the expansion of quality level ranges during training, progressively broadening the sampling range of $\Delta t$ across epochs. Performance peaks at $\alpha = 0.9$ (CT: 43.38 dB, MR: 36.16 dB), as this rate enables the model to first learn local relative flows within narrow quality ranges, then gradually extend to broader ranges, facilitating the composition of relative flows into the unified absolute flow. When $\alpha = 1.0$ (no expansion), the model learns only fixed-range flows, limiting generalization across quality levels (CT: 40.85 dB, MR: 34.83 dB). When $\alpha = 1.1 > 1$ (contraction), the quality range shrinks over training, preventing the model from learning comprehensive flow dynamics (CT: 36.39 dB, MR: 31.47 dB). The sampling time step schedule $\{\Delta t_i\}$ determines the iterative denoising trajectory, with [0.2,0.1,0.05] achieving optimal performance (CT: 43.38 dB, MR: 36.16 dB) through coarse-to-fine refinement that progressively transports images toward higher quality levels. Uniform steps [0.1,0.1,0.1] yield suboptimal results (CT: 42.93 dB, MR: 37.89 dB), while reversed schedules [0.05,0.1,0.2] significantly degrade performance (CT: 40.45 dB, MR: 36.54 dB), confirming the importance of progressive refinement in flow-based denoising.

\subsection{Trajectory Analysis}

We conduct trajectory analysis comparing DDIM~\cite{song2020ddim}, Flow Matching~\cite{lipman2022flow}, and RelativeFlow on a synthetic distribution-transport problem, where noisy references are modeled as blurred Gaussian distributions and the denoising target is a sharp circular ring distribution. This controlled setting isolates each method's transport capability from modality-specific factors, directly revealing how each framework handles the noisy reference problem.

As shown in Fig.~\ref{fig:trajectory_analysis}, RelativeFlow's iterative results are substantially closer to the clean target distribution compared to DDIM and Flow Matching. Both DDIM and Flow Matching treat the blurred Gaussian references as clean generation endpoints, causing their velocity fields to point toward heterogeneous reference distributions rather than the true clean target. This reference bias manifests as persistent cluster-like artifacts in intermediate denoising steps, preventing both baselines from recovering the circular ring structure. In contrast, RelativeFlow introduces the ring pattern markedly earlier in the transport trajectory: CoT constrains each relative flow to lie on the shared absolute flow path, ensuring structural consistency across all intermediate states, while SVF orients each denoising step along the true degradation axis rather than toward the noisy reference endpoint. Together, CoT and SVF produce a transport trajectory that converges faithfully toward the clean target throughout every step, providing geometric-level evidence for the quantitative improvements observed in Tab.~\ref{tab:combined_quant}.

\section{Conclusion}
\label{sec:Conclusion}

In this paper, we reformulate medical image denoising as learning relative flows from heterogeneous noisy references, providing a fundamental solution to the \textit{noisy reference} problem that limits existing learning paradigms. Our proposed \textbf{RelativeFlow} framework learns from noisy references at varying quality levels and drives inputs toward a unified high-quality target, achieving consistent denoising across different acquisition conditions. This is realized through two complementary components: Consistent Transport (CoT), which ensures relative flows from heterogeneous noisy references compose into a unified absolute flow, and Simulation-based Velocity Field (SVF), which constructs physics-consistent supervision via modality-specific degradation operators.
It outperforms existing SimSDL, SSL, and SimSGL methods on both CT and MR denoising tasks. 
Beyond advancing medical image denoising, RelativeFlow opens new frontiers in learning from heterogeneous noisy data, offering insights for broader denoising tasks. 
Important future work involves scaling to larger clinical datasets and extending to 3D data and other image modalities, fostering the development of robust generative models for medical imaging applications.

RelativeFlow's extrapolation beyond noisy reference quality relies on the accuracy of SVF's physics-based degradation modeling. When input noise patterns fall outside the operator's scope, residual structural artifacts may arise. The fixed inference step schedule performs reliably under standard evaluation settings but may require adaptive adjustment for more heterogeneous or severely degraded clinical inputs.

\section*{Acknowledgment}
This work was supported in part by the National Natural Science Foundation of China under Grant T2225025 (Y.C.); in part by the National Key Project of Research and Development Plan under Grant 2022YFC2401600 (Y.C.); in part by the Interdisciplinary Research Program for Young Scholars (2024FGC1004) from Southeast University (R.G.) and the Fundamental Research Funds for the Central Universities (2242025F10004) (R.G.). This research work is supported by the Big Data Computing Center of Southeast University.

{
    \small
    \bibliographystyle{ieeenat_fullname}
    \bibliography{main}
}


\newpage

\appendix

\begin{center}
{\Large \textbf{Supplementary Materials}}
\end{center}

\vspace{0.5cm}

\begin{itemize}
\item[\textbf{A.}] \hyperref[appendix:mathematical-derivations]{\textbf{Mathematical Derivations of RelativeFlow Framework}}
    \begin{itemize}
    \item[A.1] \hyperref[appendix:cot-component]{CoT Component Property}
    \item[A.2] \hyperref[appendix:cot-composition]{CoT Composition Property}
    \item[A.3] \hyperref[appendix:svf-construction]{Simulation-based Velocity Field Construction}
        \begin{itemize}
        \item[A.3.1] \hyperref[appendix:svf-flow]{Velocity Field from CoT Flow}
        \item[A.3.2] \hyperref[appendix:svf-observable]{Expressing Velocity in Terms of Observable Pairs}
        \end{itemize}
    \end{itemize}
\item[\textbf{B.}] \hyperref[appendix:experimental-details]{\textbf{Experimental Details}}
    \begin{itemize}
    \item[B.1] \hyperref[appendix:metrics]{Evaluation Metrics}
    \item[B.2] \hyperref[appendix:baselines]{Baseline Methods}
    \end{itemize}
\item[\textbf{C.}] \hyperref[appendix:dataset-details]{\textbf{Dataset Details}}
    \begin{itemize}
    \item[C.1] \hyperref[appendix:dataset-description]{Datasets for CT and MR Denoising}
    \item[C.2] \hyperref[appendix:degradation-operators]{Degradation Operators and Simulation Parameters}
    \end{itemize}
\item[\textbf{D.}] \hyperref[appendix:more-denoising-process]{\textbf{More Denoising Process Visualization}}
\end{itemize}

\vspace{0.5cm}

\section{Mathematical Derivations of RelativeFlow}
\label{appendix:mathematical-derivations}

This appendix provides the detailed mathematical derivations for the RelativeFlow framework, including the two key properties of Consistent Transport (CoT) and the Simulation-based Velocity Field (SVF) construction.

\subsection{CoT Component Property}
\label{appendix:cot-component}

We prove that any distribution in a relative flow between $p_{t_i}$ and $p_{t_j}$ is also a component of the absolute flow between $p_0$ and $p_{+\infty}$.

Consider arbitrary quality levels $0 < t_i < t < t_j < +\infty$. Both endpoints lie on the absolute flow path:
\begin{align}
p_{t_i} &= e^{-t_i} p_0 + (1-e^{-t_i}) p_{+\infty} \\
p_{t_j} &= e^{-t_j} p_0 + (1-e^{-t_j}) p_{+\infty}
\end{align}

According to the CoT probability path definition (Eq.~\eqref{eq:relative_flow}), any intermediate distribution $p_t$ for $t \in (t_i, t_j)$ is given by:
\begin{equation}
p_t = \lambda \, p_{t_i} + (1 - \lambda) \, p_{t_j}, \quad \lambda = \frac{e^{-t} - e^{-t_j}}{e^{-t_i} - e^{-t_j}}
\end{equation}

Substituting the expressions for $p_{t_i}$ and $p_{t_j}$:
\begin{align}
p_t &= \lambda \left(e^{-t_i} p_0 + (1-e^{-t_i}) p_{+\infty}\right) \notag \\
&\quad + (1-\lambda) \left(e^{-t_j} p_0 + (1-e^{-t_j}) p_{+\infty}\right) \notag \\
&= \left(\lambda e^{-t_i} + (1-\lambda) e^{-t_j}\right) p_0 \notag \\
&\quad + \left(\lambda (1-e^{-t_i}) + (1-\lambda)(1-e^{-t_j})\right) p_{+\infty}
\end{align}

Computing the coefficient of $p_0$:
\begin{align}
&\lambda e^{-t_i} + (1-\lambda) e^{-t_j} \notag \\
&= \frac{e^{-t} - e^{-t_j}}{e^{-t_i} - e^{-t_j}} e^{-t_i} + \frac{e^{-t_i} - e^{-t}}{e^{-t_i} - e^{-t_j}} e^{-t_j} \notag \\
&= \frac{e^{-t_i}e^{-t} - e^{-t_i}e^{-t_j} + e^{-t_i}e^{-t_j} - e^{-t}e^{-t_j}}{e^{-t_i} - e^{-t_j}} \notag \\
&= \frac{e^{-t}(e^{-t_i} - e^{-t_j})}{e^{-t_i} - e^{-t_j}} = e^{-t}
\end{align}

Similarly, the coefficient of $p_{+\infty}$ equals:
\begin{equation}
\lambda (1-e^{-t_i}) + (1-\lambda)(1-e^{-t_j}) = 1 - e^{-t}
\end{equation}

Therefore:
\begin{equation}
p_t = e^{-t} p_0 + (1-e^{-t}) p_{+\infty}
\end{equation}

This exactly matches the absolute flow formula, confirming that any distribution in a relative flow is a component of the absolute flow.

\subsection{CoT Composition Property}
\label{appendix:cot-composition}

We prove that for any three quality levels $0 < t_1 < t_2 < t_3$, composing flows $\psi_{t_1 \to t_2}$ and $\psi_{t_2 \to t_3}$ yields the direct flow $\psi_{t_1 \to t_3}$.

\textbf{Step 1: Express $p_{t_2}$ from flow $\psi_{t_1 \to t_2}$.}
For the relative flow from $t_1$ to $t_2$, any distribution $p_t$ at time $t \in [t_1, t_2]$ follows:
\begin{equation}
p_t = \lambda_{12} p_{t_1} + (1-\lambda_{12}) p_{t_2}, \quad \lambda_{12} = \frac{e^{-t} - e^{-t_2}}{e^{-t_1} - e^{-t_2}}
\end{equation}

Rearranging to solve for $p_{t_2}$:
\begin{align}
p_{t_2} &= \frac{p_t - \lambda_{12} p_{t_1}}{1 - \lambda_{12}} \notag \\
&= \frac{p_t - \frac{e^{-t} - e^{-t_2}}{e^{-t_1} - e^{-t_2}} p_{t_1}}{1 - \frac{e^{-t} - e^{-t_2}}{e^{-t_1} - e^{-t_2}}} \notag \\
&= \frac{e^{-t_1} - e^{-t_2}}{e^{-t_1} - e^{-t}} p_t - \frac{e^{-t} - e^{-t_2}}{e^{-t_1} - e^{-t}} p_{t_1}.
\label{eq:appendix_pt2_from_t1}
\end{align}

\textbf{Step 2: Express $p_{t_2}$ from flow $\psi_{t_2 \to t_3}$.}
For the relative flow from $t_2$ to $t_3$, we have:
\begin{equation}
p_t = \lambda_{23} p_{t_2} + (1-\lambda_{23}) p_{t_3}, \quad \lambda_{23} = \frac{e^{-t} - e^{-t_3}}{e^{-t_2} - e^{-t_3}}
\end{equation}

Rearranging to solve for $p_{t_2}$:
\begin{align}
p_{t_2} &= \frac{p_t - (1-\lambda_{23}) p_{t_3}}{\lambda_{23}} \notag \\
&= \frac{p_t - \left(1 - \frac{e^{-t} - e^{-t_3}}{e^{-t_2} - e^{-t_3}}\right) p_{t_3}}{\frac{e^{-t} - e^{-t_3}}{e^{-t_2} - e^{-t_3}}} \notag \\
&= \frac{e^{-t_2} - e^{-t_3}}{e^{-t} - e^{-t_3}} p_t - \frac{e^{-t_2} - e^{-t}}{e^{-t} - e^{-t_3}} p_{t_3}.
\label{eq:appendix_pt2_from_t3}
\end{align}

\textbf{Step 3: Eliminate $p_{t_2}$ to obtain the composed flow.}
Equating Eq.~\eqref{eq:appendix_pt2_from_t1} and Eq.~\eqref{eq:appendix_pt2_from_t3}:
\begin{align}
&\frac{e^{-t_1} - e^{-t_2}}{e^{-t_1} - e^{-t}} p_t - \frac{e^{-t} - e^{-t_2}}{e^{-t_1} - e^{-t}} p_{t_1} \notag \\
&\quad = \frac{e^{-t_2} - e^{-t_3}}{e^{-t} - e^{-t_3}} p_t - \frac{e^{-t_2} - e^{-t}}{e^{-t} - e^{-t_3}} p_{t_3}.
\end{align}

Collecting terms with $p_t$:
\begin{align}
&\left(\frac{e^{-t_1} - e^{-t_2}}{e^{-t_1} - e^{-t}} - \frac{e^{-t_2} - e^{-t_3}}{e^{-t} - e^{-t_3}}\right) p_t \notag \\
&\quad = \frac{e^{-t} - e^{-t_2}}{e^{-t_1} - e^{-t}} p_{t_1} - \frac{e^{-t_2} - e^{-t}}{e^{-t} - e^{-t_3}} p_{t_3}.
\end{align}

After algebraic simplification (omitted for brevity), we obtain:
\begin{equation}
p_t = \frac{e^{-t} - e^{-t_3}}{e^{-t_1} - e^{-t_3}} p_{t_1} + \frac{e^{-t_1} - e^{-t}}{e^{-t_1} - e^{-t_3}} p_{t_3}.
\end{equation}

This exactly matches the CoT probability path for the direct flow from $t_1$ to $t_3$ with $\lambda_{13} = \frac{e^{-t} - e^{-t_3}}{e^{-t_1} - e^{-t_3}}$, confirming the composition property $\psi_{t_1 \to t_3} = \psi_{t_2 \to t_3} \circ \psi_{t_1 \to t_2}$.

\subsection{Simulation-based Velocity Field Construction}
\label{appendix:svf-construction}

This subsection provides the detailed mathematical derivation showing how the velocity field can be constructed from observable pairs $(x_{t-\Delta t}, x_t)$ without requiring access to clean endpoints or absolute time labels.

\subsubsection{Velocity Field from CoT Flow}
\label{appendix:svf-flow}

Following the CoT-defined flow matching framework, let $\psi_t$ denote the flow map that transports from the noise endpoint at $t=0$ to quality level $t$. For a sample $x_0$ drawn from the noise distribution $p_0$, the CoT path yields:
\begin{equation}
\psi_t(x_0) = e^{-t} x_0 + (1-e^{-t}) x_{\infty}
\label{eq:cot_flow}
\end{equation}
where $x_{\infty}$ represents the clean endpoint as $t \to +\infty$.

The velocity field is defined as the time derivative of the flow:
\begin{align}
u_t\big(\psi_t(x_0)\,|\,x_0, x_{\infty}\big) &= \frac{d}{dt}\, \psi_t(x_0) \notag \\
&= -e^{-t} x_0 + e^{-t} x_{\infty} \\
& = e^{-t}(x_{\infty}-x_0).
\label{eq:velocity_absolute}
\end{align}

\subsubsection{Expressing Velocity in Terms of Observable Pairs}
\label{appendix:svf-observable}

In practice, we only observe a noisy reference $x_t$ at arbitrary quality level $t$ and its degraded version $x_{t-\Delta t} = D_{\Delta t}(x_t)$ at an earlier quality level. Both samples lie on the same absolute flow path, so they can be expressed using Eq.~\eqref{eq:cot_flow}:
\begin{align}
x_t &= e^{-t} x_0 + (1-e^{-t}) x_{\infty} \\
x_{t-\Delta t} &= e^{-(t-\Delta t)} x_0 + (1-e^{-(t-\Delta t)}) x_{\infty}
\end{align}

Subtracting the second equation from the first:
\begin{align}
x_t - x_{t-\Delta t} &= \left(e^{-t} - e^{-(t-\Delta t)}\right) x_0 \notag \\
&\quad + \left((1-e^{-t}) - (1-e^{-(t-\Delta t)})\right) x_{\infty} \notag \\
&= \left(e^{-t} - e^{-t}e^{\Delta t}\right) x_0 + \left(e^{-t}e^{\Delta t} - e^{-t}\right) x_{\infty} \notag \\
&= e^{-t}(1 - e^{\Delta t}) x_0 + e^{-t}(e^{\Delta t} - 1) x_{\infty} \notag \\
&= e^{-t}(e^{\Delta t} - 1) (x_{\infty} - x_0)
\end{align}

Solving for $(x_{\infty} - x_0)$:
\begin{equation}
x_{\infty} - x_0 = \frac{x_t - x_{t-\Delta t}}{e^{-t}(e^{\Delta t} - 1)} = \frac{x_t - x_{t-\Delta t}}{e^{-(t-\Delta t)} - e^{-t}}
\end{equation}

Substituting this into the velocity field expression from Eq.~\eqref{eq:velocity_absolute}:
\begin{align}
u_{t}(x_{t}) &= e^{-t}(x_{\infty}-x_0) \notag \\
&= e^{-t} \cdot \frac{x_t - x_{t-\Delta t}}{e^{-t}(e^{\Delta t} - 1)} \notag \\
&= \frac{x_t - x_{t-\Delta t}}{e^{\Delta t} - 1}
\end{align}

Note that by the CoT component property (Appendix~\ref{appendix:cot-component}), the relative flow between $x_{t-\Delta t}$ and $x_t$ follows the same CoT form as the absolute flow. Therefore, the above velocity $u_t(x_t)$ at the absolute quality level $t$ corresponds to the velocity field in the relative flow with parameter $\Delta t$. For notational clarity in the training objective, we rewrite this velocity as a function of the step size $\Delta t$ and the observable pair:
\begin{equation}
u_{\Delta t}(x_{t-\Delta t}|x_t) = \frac{x_t - x_{t-\Delta t}}{e^{\Delta t}-1}
\label{eq:normalized_velocity}
\end{equation}

This expression matches the supervision target in Eq.~\eqref{eq:training_loss}, showing that the velocity field can be computed directly from the observable degradation pair $(x_{t-\Delta t}, x_t)$ without requiring knowledge of the absolute endpoints $(x_0, x_{\infty})$ or the absolute quality level $t$.

\begin{figure*}[!t]
\centering
\includegraphics[width=0.95\textwidth]{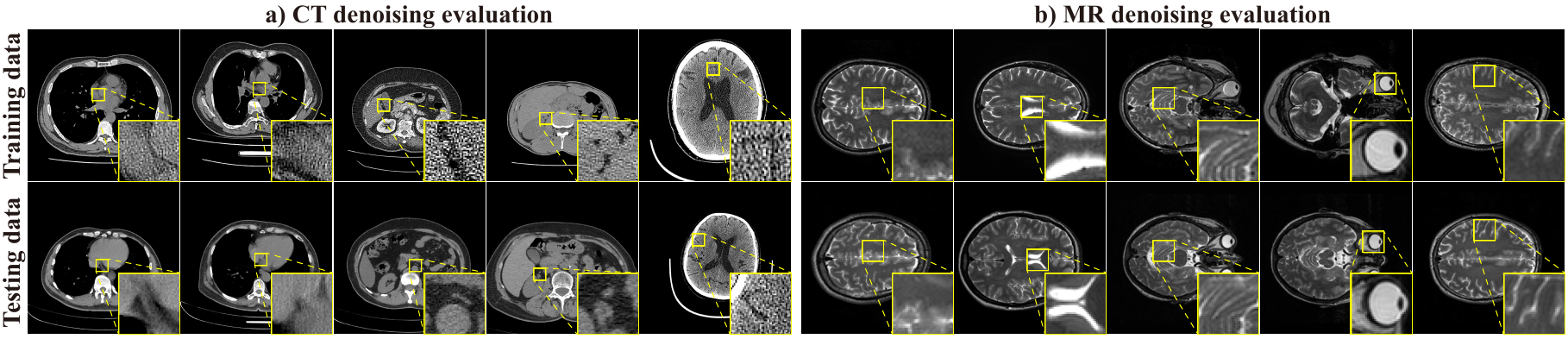}
\caption{Visual comparison of training and testing data quality distributions for CT (left) and MR (right) datasets. \textbf{Top row:} Training data exhibit heterogeneous quality levels with varying degrees of degradation across different acquisition protocols, exemplifying the noisy reference problem. \textbf{Bottom row:} Testing data were manually selected to include only the highest-quality reference images for reliable evaluation.}
\label{fig:dataset-distribution}
\end{figure*}

\section{Experimental Details}
\label{appendix:experimental-details}

\subsection{Evaluation Metrics}
\label{appendix:metrics}

We evaluate denoising performance using four standard metrics following the main text. Images are normalized to $[0,1]$ range for computing PSNR, SSIM, and RMSE, and to 3-channel $[0,255]$ format for LPIPS computation. For clearer numerical comparison, SSIM, RMSE, and LPIPS are reported as their raw scores multiplied by 100.

\textbf{Peak Signal-to-Noise Ratio (PSNR):}
\begin{equation}
\text{PSNR} = 10 \log_{10} \frac{1}{\text{MSE}}
\end{equation}
where $\text{MSE} = \frac{1}{N}\sum_{i=1}^{N}(x_i - y_i)^2$ is the mean squared error between the denoised image $x$ and ground truth $y$ in $[0,1]$ range, and $N$ is the total number of pixels.

\textbf{Structural Similarity Index Measure (SSIM):}
\begin{equation}
\text{SSIM}(x, y) = \frac{(2\mu_x\mu_y + c_1)(2\sigma_{xy} + c_2)}{(\mu_x^2 + \mu_y^2 + c_1)(\sigma_x^2 + \sigma_y^2 + c_2)}
\end{equation}
where $\mu_x$, $\mu_y$ are local means, $\sigma_x^2$, $\sigma_y^2$ are local variances, $\sigma_{xy}$ is local covariance, and $c_1 = (0.01)^2$, $c_2 = (0.03)^2$ are stabilizing constants.

\textbf{Normalized Root Mean Squared Error (RMSE):}
\begin{equation}
\text{RMSE} = \sqrt{\frac{1}{N}\sum_{i=1}^{N}(x_i - y_i)^2}
\end{equation}
This is computed on $[0,1]$ normalized images. Lower values indicate better performance.

\textbf{Learned Perceptual Image Patch Similarity (LPIPS):}
\begin{equation}
\text{LPIPS}(x, y) = \sum_{l} w_l \cdot \|\phi_l(x) - \phi_l(y)\|_2^2
\end{equation}
where $\phi_l$ represents features from layer $l$ of a pretrained VGG network, and $w_l$ are learned weights. Images are converted to 3-channel $[0,255]$ format before computing LPIPS. Lower values indicate better perceptual similarity.

\subsection{Baseline Methods}
\label{appendix:baselines}

We compare RelativeFlow with 10 baseline methods from three categories under the unified settings summarized in Table~\ref{tab:unified-config}. 
 
\begin{table}[h]
\centering
\caption{Unified training configuration for all baseline methods.}
\label{tab:unified-config}
\resizebox{\columnwidth}{!}{
\begin{tabular}{lll}
\toprule
Category & Parameter & Setting \\
\midrule
\multirow{5}{*}{Training} & Loss function & L2 loss \\
& Optimizer & Adam \\
& Learning rate & $10^{-4}$ \\
& Optimizer betas & $\beta_1=0.9$, $\beta_2=0.999$ \\
& Training epochs & 30 \\
& Inference steps (SimSGL) & 3 \\
\midrule
Hardware & GPUs & 8 $\times$ NVIDIA V100 (32 GB) \\
\midrule
\multirow{2}{*}{Architecture} & SimSDL/SSL & 2D U-Net~\cite{ronneberger2015u} \\
& SimSGL & 2D Guided U-Net~\cite{dhariwal2021diffusion} \\
\bottomrule
\end{tabular}
}
\end{table}

For SimSDL methods, SwinIR~\cite{liang2021swinir} replaces standard convolution with window-based self-attention and shifted window mechanism for hierarchical feature learning. RED-CNN~\cite{chen2017redcnn} uses symmetric encoder-decoder architecture with shortcut connections between corresponding layers. CNN-DMRI~\cite{tripathi2020cnn} incorporates k-space consistency enforcement to ensure MR reconstruction matches observed measurements.

For SSL methods, Noise2Self~\cite{batson2019noise2self} implements blind-spot networks where each pixel is predicted from spatially disjoint context via donut masking. Noise2Sim~\cite{niu2022noise2sim} exploits patch redundancy by identifying and averaging structurally similar patches within single images. Deformed2Self~\cite{xu2021deformed2self} enforces deformation consistency through spatial transformer networks.

For SimSGL methods, DDIM~\cite{song2020ddim} uses non-Markovian diffusion process for deterministic sampling with $T=1000$ training steps. Flow Matching~\cite{lipman2022flow} directly regresses velocity fields via optimal transport paths. IPDM~\cite{liao2024idpm} applies partial diffusion starting from noisy observations rather than pure noise. R2D2+~\cite{chung2022r2d2} performs posterior sampling with predictor-corrector strategy for inverse problems.

\begin{figure*}[!htbp]
\centering
\includegraphics[width=\linewidth]{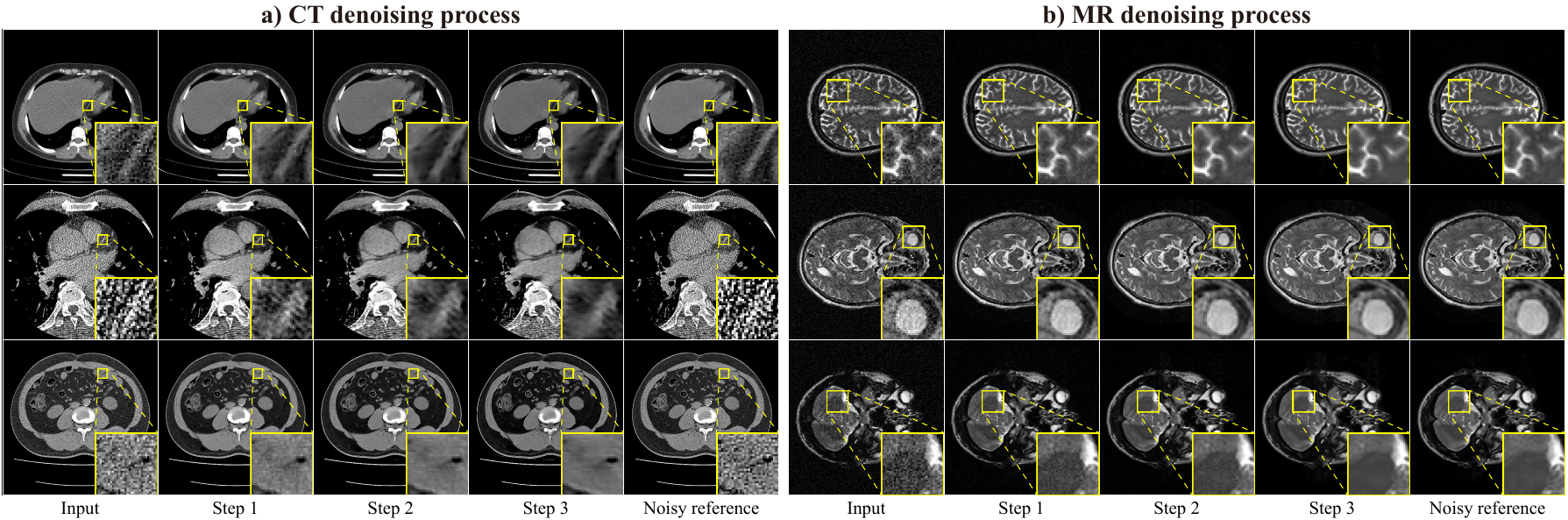}
\caption{\textbf{Denoising process visualization for multiple quality levels.} Left: three CT examples. Right: three MR examples. Each row shows the denoising trajectory with five images: noisy input, three intermediate steps, and the reference. Yellow boxes indicate ROI with zoomed-in views. RelativeFlow achieves consistent high-quality outputs across different noise levels and modalities.}
\label{fig:denoising_process_more}
\end{figure*}

\section{Dataset Details}
\label{appendix:dataset-details}

\subsection{Datasets for CT and MR Denoising}
\label{appendix:dataset-description}

\begin{table}[h]
\centering
\caption{Dataset statistics for training, testing, and validation splits (patients / slices).}
\label{tab:dataset-statistics}
\begin{tabular}{lcc}
\toprule
Split & GBA-LDCT dataset & IXI dataset\\
\midrule
Training & 263 / 138,440 & 497 / 129,566 \\
Testing & 27 / 12,342 & 40 / 10,360 \\
Validation & 27 / 18,856 & 42 / 11,064 \\
\bottomrule
\end{tabular}
\end{table}

Both datasets naturally exhibit the noisy reference problem with heterogeneous quality levels across acquisition protocols, as illustrated in Figure~\ref{fig:dataset-distribution}. 
For the \textbf{training data} (top row), the GBA-LDCT dataset used in CT denoising evaluation contains images acquired from different manufacturers and anatomical regions with varying scanning protocols, including different tube voltages, tube currents, and reconstruction kernels, resulting in substantial quality variations across samples. 
Similarly, the IXI dataset used in MR denoising evaluation includes images collected from multiple medical centers using different manufacturers' scanners with varying field strengths, pulse sequences, and acquisition parameters, leading to heterogeneous image quality levels. 
These quality variations naturally present in clinical data exemplify the noisy reference problem addressed in our work.

In contrast, the \textbf{testing data} (bottom row) for both datasets were manually selected to include only the highest-quality reference images, ensuring reliable evaluation of denoising performance. 
The complete dataset statistics are summarized in Table~\ref{tab:dataset-statistics}.

\subsection{Degradation Operators and Simulation Parameters}
\label{appendix:degradation-operators}

\subsubsection{CT Degradation Operator}

For CT images, the degradation operator models quantum noise and electronic noise following Poisson-Gaussian distribution. In the sinogram domain where $x$ represents normalized measurements, we define:
\begin{equation}
D_{\Delta t}(x) = \frac{\text{Poisson}(\alpha(\Delta t) \cdot I_0 \cdot x) + \epsilon(\Delta t)}{\alpha(\Delta t) \cdot I_0}
\end{equation}
where $\alpha(\Delta t) = \exp(-\lambda_{\text{CT}} \Delta t)$ models dose reduction, $I_0$ is the incident photon count, and $\epsilon(\Delta t) \sim \mathcal{N}(0, \sigma_e^2(\Delta t))$ represents electronic noise with $\sigma_e^2(\Delta t) = \beta_{\text{CT}} \Delta t$.

Parameters: $\lambda_{\text{CT}} = 5$, $\beta_{\text{CT}} = 400$, $I_0 = 10^6$ photons.

\subsubsection{MR Degradation Operator}

For MR images, the degradation operator models thermal noise and dielectric losses following Rician distribution in the magnitude domain:
\begin{equation}
D_{\Delta t}(x) = \sqrt{(x + n_1(\Delta t))^2 + n_2^2(\Delta t)}
\end{equation}
where $n_1(\Delta t), n_2(\Delta t) \sim \mathcal{N}(0, \sigma^2(\Delta t))$ are independent Gaussian noise components with variance $\sigma^2(\Delta t) = \gamma_{\text{MR}} \Delta t$.

Parameters: $\gamma_{\text{MR}} = 0.005$.

\section{Denoising Process Visualization}
\label{appendix:denoising-process}

To further demonstrate the effectiveness of RelativeFlow, we provide additional visualizations of the iterative denoising process on both CT and MR images with varying noise levels. 
As shown in Figure~\ref{fig:denoising_process_more}, for inputs with different noise levels, RelativeFlow progressively improves the image quality through three intermediate denoising steps, ultimately achieving consistently high-quality results that closely match the reference images. 
Each row displays the complete denoising trajectory: from the noisy input (left), through three successive refinement steps, to the final high-quality reference (right). 
The yellow-highlighted regions of interest (ROI) and their zoomed-in views clearly illustrate the progressive noise reduction and detail preservation across different modalities and quality levels.

\end{document}